\documentclass[%
 aip,
 amsmath,amssymb,
 reprint,%
]{revtex4-1}

\usepackage{graphicx}
\usepackage{dcolumn}
\usepackage{bm}

\usepackage{xcolor}

\usepackage[utf8]{inputenc}
\usepackage[T1]{fontenc}
\usepackage{mathptmx}
\usepackage{etoolbox}

\usepackage{braket}
\usepackage{upgreek}
\usepackage{siunitx}
\usepackage{booktabs}

\makeatletter
\def\@email#1#2{%
 \endgroup
 \patchcmd{\titleblock@produce}
  {\frontmatter@RRAPformat}
  {\frontmatter@RRAPformat{\produce@RRAP{*#1\href{mailto:#2}{#2}}}\frontmatter@RRAPformat}
  {}{}
}%
\makeatother
\begin{document}

\preprint{AIP/123-QED}

\title{Superconducting Quantum Memory with a Suspended Coaxial Resonator}
\author{Lev Krayzman}
 \email{lev.krayzman@princeton.edu}
\thanks{Currently at Princeton University}
\author{Chan U Lei}
\email{lei@quantumcircuits.com}
\thanks{Currently at Quantum Circuits, Inc.}
\author{Suhas Ganjam}
\author{James Teoh}
\thanks{Currently at Quantum Circuits, Inc.}
\author{Luigi Frunzio}
\author{Robert J. Schoelkopf}
\affiliation{Department of Applied Physics, Yale University, New Haven, Connecticut 06511, USA}
\affiliation{Yale Quantum Institute, Yale University, New Haven, Connecticut 06520, USA}

\date{\today}

\begin{abstract}
A promising way to store quantum information is by encoding it in the bosonic excitations of microwave resonators.
This provides for long coherence times, low dephasing rates, as well as a hardware-efficient approach to quantum error correction.
There are two main methods used to make superconducting microwave resonators: traditionally machined out of bulk material, and lithographically fabricated on-chip in thin film.
3D resonators have few loss channels and larger mode volumes, and therefore smaller participations in the lossy parts, but it can be challenging to reach high material qualities.
On-chip resonators can use low-loss thin films, but confine the field more tightly, resulting in higher participations and additional loss channels from the dielectric substrate. In this work, we present a design in which a dielectric scaffold supports a thin-film conductor within a 3D package, thus combining the low surface participations of bulk-machined cavities with the high quality and control over materials of thin-film circuits. 
By incorporating a separate chip containing a transmon qubit, we realize a quantum memory and measure single-photon lifetimes in excess of a millisecond. This hybrid 3D architecture has several advantages for scaling, as it relaxes the importance of the package and permits modular construction with separately-replaceable qubit and resonator devices.
\end{abstract}

\maketitle

Superconducting circuits are one of the leading platforms for quantum computing.
One of the ways of using them to build a quantum computer is to store information in microwave-frequency linear resonators with bosonic codes \cite{joshi_quantum_2021}.
This provides the advantages of a large Hilbert space to store information, which enables hardware efficiency in storing a logical qubit in a small number of physical elements, as well as the presence of only one dominant error channel, photon loss.
The lifetimes of the microwave resonators used to store quantum information have been increasing, and have reached the tens of milliseconds when coupled to a nonlinear ancilla \cite{milul_superconducting_2023} and even longer on their own \cite{romanenko_three-dimensional_2020}.

There are two primary ways that these resonators are made: traditionally machined from bulk superconductor, and lithographically fabricated from thin films on a chip.
3D cavities have a large mode volume compared to their on-chip counterparts, which dilutes the field inside of lossy materials, such as metal surfaces.
They also do not require a lossy dielectric substrate. 
These features allow them to achieve long lifetimes, with the most common design of the coaxial stub cavity \cite{reagor_quantum_2016} routinely achieving millisecond coherence when coupled to a nonlinear ancilla, and newer designs going to the tens of milliseconds or beyond\cite{milul_superconducting_2023, romanenko_three-dimensional_2020}.
However, in order to achieve high quality, the cavities must be made of materials that are difficult to process such as high-purity aluminum or niobium, that are then treated in complicated ways such as acid etching, vacuum baking, and welding \cite{chakram_seamless_2021,kudra_high_2020,heidler_non-markovian_2021}.
These difficulties, as well as challenges in scaling up the numbers of traditionally-machined cavities, have resulted in the fact that no device to date has yet exceeded a handful of cavities with individual quantum control \cite{Chou_Blumoff_Wang_Reinhold_Axline_Gao_Frunzio_Devoret_Jiang_Schoelkopf_2018, zhou_realizing_2023}.

On the other hand, on-chip devices use standard lithographic device fabrication, and are thus relatively straightforward to scale up in number.
Over the past several years, their coherences have also been improved, with Q's of millions to over ten million at single-photon powers becoming attainable \cite{altoe_localization_2022, crowley_disentangling_2023, deng_titanium_2023, ganjam_surpassing_2023}.
However, the on-chip designs tend to concentrate the field in several potentially-lossy regions, and thus still lag behind the 3D cavities in terms of lifetime.

It is therefore desirable to combine the advantages of the two approaches to create more scalable high-Q resonators.
One approach is to micromachine a cavity into silicon and cover it in a superconducting thin film, which uses techniques learned from classical circuits and silicon fabrication to make a superconducting 3D cavity \cite{brecht_multilayer_2016, brecht_micromachined_2017-1}.
Despite the increased surface participations due to smaller dimensions, the micromachined cavity was able to achieve higher lifetimes than the coaxial stub cavity \cite{lei_high_2020} due to the higher quality of its thin-film materials \cite{lei_characterization_2023}.
However, the relative difficulty of the fabrication and the unresolved issue of inter-cavity coupling pose  barriers to its utilization.

In this work, we introduce an alternate superconducting microwave resonator quantum memory architecture: a suspended coaxial resonator comprising a thin-film conductor surrounding a dielectric scaffold, housed in a machined 3D package. 
With this approach, we eliminate the contribution to the loss from the dielectric substrate and dilute the fields, thereby decreasing the participations in other lossy materials such as the conductor and metal-air interfaces.
Additionally, the resonator remains very similar to the commonly-used stub cavity, making it easier to use and providing a known path to couple it to non-linear elements, which is demonstrated by combining the resonator with a transmon on a 
separate chip. 

\begin{figure*}
\centering
\includegraphics[width=0.8\textwidth]{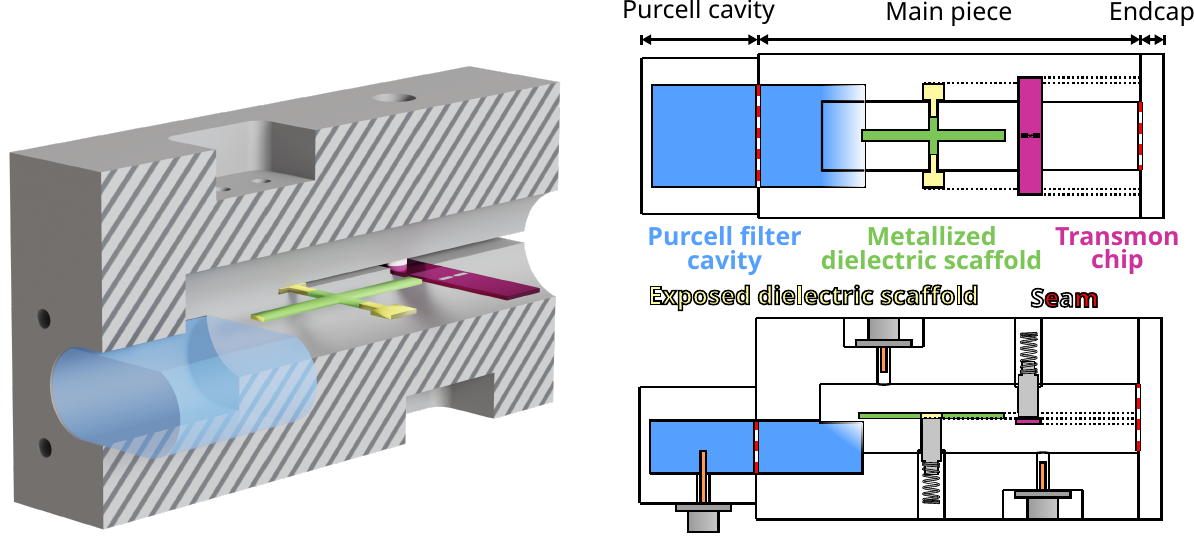}
\caption{Left: a 3D render of the main piece of the device, with the outer conductor sectioned in half for visibility \textendash{} note that the outer conductor is actually machined as one piece and is not cut in half.
Right: top and side view diagrams of the device.
The Purcell filter cavity is shown in blue, the bare cross-shaped centerpin chip is in yellow with the metallization on it in green, and the transmon chip is in magenta.
The full package consists of three pieces: the main piece, as shown in gray on the left, a second piece containing the other half of the Purcell filter cavity, and an endcap.
The two seams between the three machined parts of the device are indicated by the dashed red-and-white line.
The three coupling pins as well as the PTFE clips and springs for the chips are visible in the side view; a PTFE clip can also be seen pressing on the transmon chip in the render on the left.
The tunnel to the endcap is sufficiently long that the seam loss at the joint is negligible (same as for a stub cavity).
For the Purcell cavity, the seam loss is much higher, but the cavity is overcoupled to the readout port, and so it does not matter.
}
\label{fig:device_diagram}
\end{figure*}

A diagram of this resonator is shown in Fig.~\ref{fig:device_diagram}.
This device comprises a $\lambda/2$ resonator, in which the center conductor is an evaporated thin film coating a dielectric support structure, and the outer conductor is a tunnel traditionally machined into a bulk superconductor.
The transmission line modes of this resonator (as well as their participations) are essentially the same as that of the coaxial stub cavity, which is instead a $\lambda/4$ resonator.
Our device allows the center conductor to be physically separated from the outer conductor, enabling the use of evaporated  thin-film materials with higher quality. 

\begin{figure}
  \centering
  \includegraphics[width=1\linewidth]{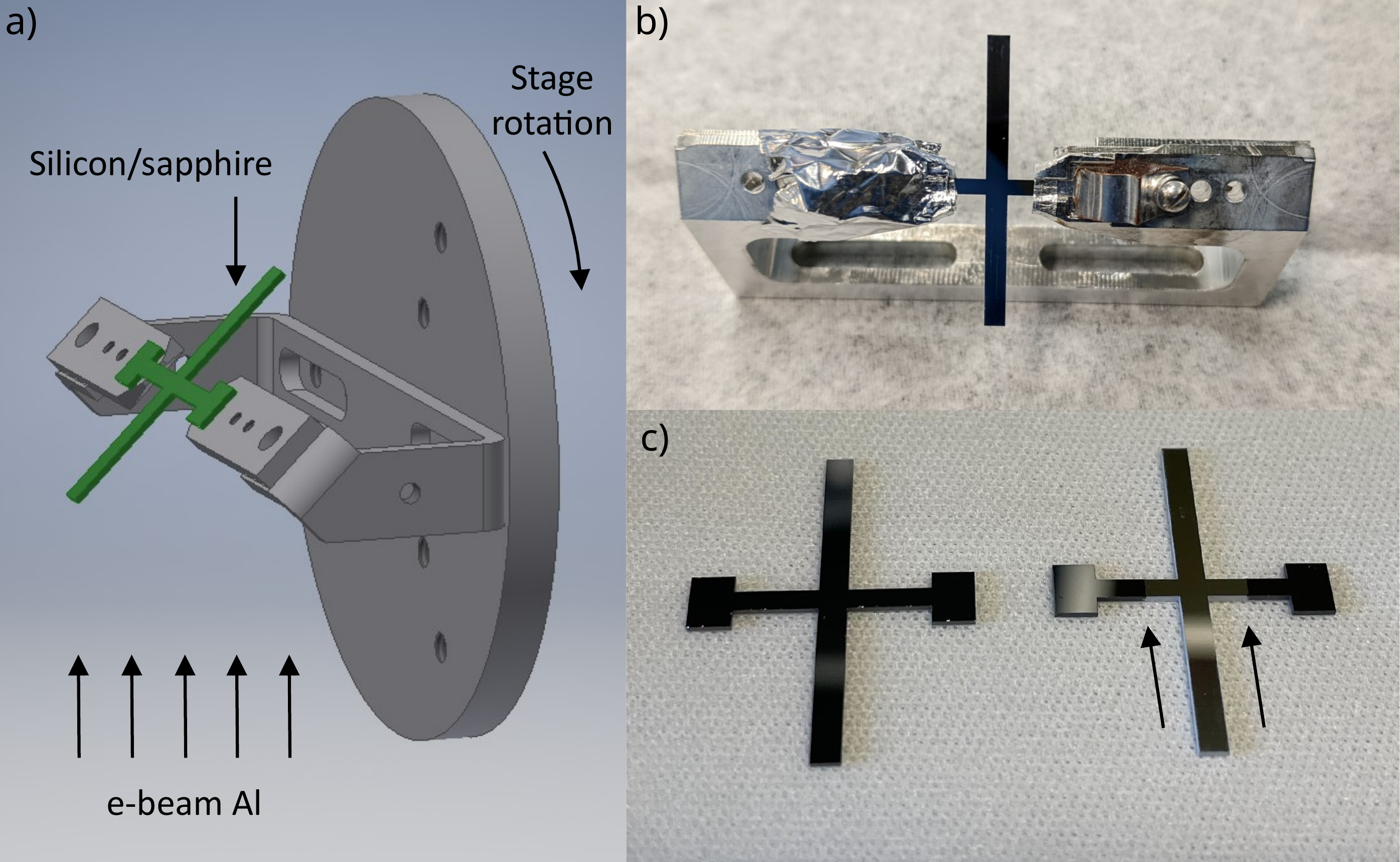}
  \caption{
  Evaporation of superconductor on all sides of the dielectric scaffold.
  a) The scaffold is held at a 45\textdegree{} angle relative to the evaporation axis and rotated perpendicularly to it.
  This provides coverage from all directions.
  b) A photograph of the clamping mechanism with a center conductor in it.
  The scaffold is held in place with a beryllium copper clamp (visible on the right), with a small piece of aluminum foil shadowing the sharp line between the coated and uncoated parts.
  The whole area is then wrapped in more foil (visible on the left) in order to shadow the entire uncoated region.
  c) A blank silicon scaffold (left), and a silicon scaffold post-evaporation (right).
  The arrows point to the extent of the metallization on the shorter mechanical support beam.
  }
  \label{fig:evaporation}
\end{figure}

The supporting dielectric scaffold is lasercut into a cross shape out of silicon or sapphire.
The longer cross beam serves as the stripline center conductor, once metallized, whereas the other exists only as an unmetallized mechanical support for clamping on both sides.
An important feature of the center conductor is that the device is rotated inside the evaporation system while the thin-film superconductor is deposited such that the dielectric is coated from all sides (most of the mechanical support beam is masked with aluminum foil), see Fig.~\ref{fig:evaporation}.
This means that the $\lambda/2$ mode, which has an electric field node at the center of the resonator, has almost no participation in the dielectric, see Table~\ref{tab:storage_participations}.
As a result, unlike on-chip stripline resonators defined lithographically on a chip, our storage mode is not limited by the lossy dielectric substrate \cite{read_precision_2023}.
The 3$\lambda$/2 mode of this resonator is used for readout.
This readout mode does have participation in the dielectric scaffold, but since it is engineered to have a low coupling $Q$, this is less important.
Readout is performed through the Purcell filter mode, which lives in a separate volume on the side of the main cavity.
This mode has a seam in the middle and thus experiences seam loss \cite{brecht_demonstration_2015}.
As with the readout mode, the Purcell filter mode is engineered to have a very low coupling $Q$, so the seam loss is not important.

\begin{table}
\caption{Most important participations of the storage mode for the 100 \textmu{}m-thick lasercut sapphire scaffold.
The participations are fractions of total mode energy stored in the given loss channel while the q's are the loss factors for that loss channel (e.g. $1/\tan\delta$ for a dielectric).
The associated loss model is $\frac{1}{Q_i} = \sum_i \frac{p_i}{q_i}$ with $p$ the participations and $i$ the loss channel; the $Q_i$ limit from channel $i$ is $q_i/p_i$.
The lasercut scaffold participations change slightly for the 500\,\textmu{}m-thick silicon, but the expected Q's are still very high.
Note that the sidewalls of the stripline are not included in this model, as their quality has not yet been separately measured.
Expected q for EFG sapphire is from Ref.~\onlinecite{read_precision_2023}, aluminum on sapphire bounds are from Ref.~\onlinecite{ganjam_improving_2023}, and bulk aluminum properties are from Ref.~\onlinecite{lei_characterization_2023}.
For the bulk aluminum, relatively conservative values are chosen; the values presented are for the 6061 alloy and for 5N aluminum etched with the standard process.
The seam loss on the last line is expressed in units of $y_\text{seam}$ and $g_\text{seam}$, both in $1/(\Omega \text{m})$; the seam quality is not significantly different between 6061 and etched 5N aluminum.} 
\setlength{\tabcolsep}{5.5pt}
    \centering
    \begin{tabular}{llrr}
        \toprule{}
         Loss channel & Participation & Expected q & $Q_i$ limit \\
         \midrule{}Lasercut chip bulk & $5\times10^{-5}$ & $1.6\times10^7$ & $3\times 10^{11}$ \\
         Lasercut chip SA & $5\times10^{-10}$ & $8.3\times10^2$ & $2\times10^{12}$ \\
         Qubit chip bulk & $1\times10^{-3}$ & $1.6\times10^7$ & $2\times10^{10}$ \\
         Stripline conductor & $2.5\times10^{-5}$ & $>2.0\times10^5$ & $>8\times10^9$\\ 
         Stripline MA & $2\times10^{-7}$ & $>1.7\times10^2$ & $>9\times10^8$\\ 
         Package conductor & $3.5\times10^{-6}$ & 400 (6061) & $1\times10^8$ \\
         & & 3000 (5N) & $9\times10^8$ \\
         Package MA & $1.5\times10^{-8}$ & 10 (6061) & $7\times10^8$ \\ 
         & & 20 (5N) & $1\times10^9$ \\
         Purcell cavity seam & $3\times10^{-7}$ & $2.5\times10^4$ & $8\times10^{10}$ \\
         \midrule{}Expected total $Q_i$ & & 6061 & $8\times10^7$ \\
         & & 5N & $3\times10^8$ \\
         \bottomrule{}
    \end{tabular}
    \label{tab:storage_participations}
\end{table}

A standard electron-beam lithography-fabricated 3D transmon on a 100\,\textmu{}m-thick sapphire chip is used as an ancilla.
The center conductor and transmon chip are inserted into special slots in the package, which are made using wire electrical discharge machining.
The center conductor dielectric scaffold and transmon chip are made to have different sizes, so that they stay in their own slots.
They are held in place by friction: PTFE rods push the dielectric against the wall of the slot.
Force is applied to the rods by beryllium copper springs, which are loaded into a tunnel and pre-compressed with  metal set screws.
This clamping design was chosen as one that eliminated vibration of the dielectric scaffold; see the Supplemental Material and Ref.~\onlinecite{Krayzman_thin-film_2022} for details.

\begin{figure}
  \centering
  \includegraphics[width=1\linewidth]{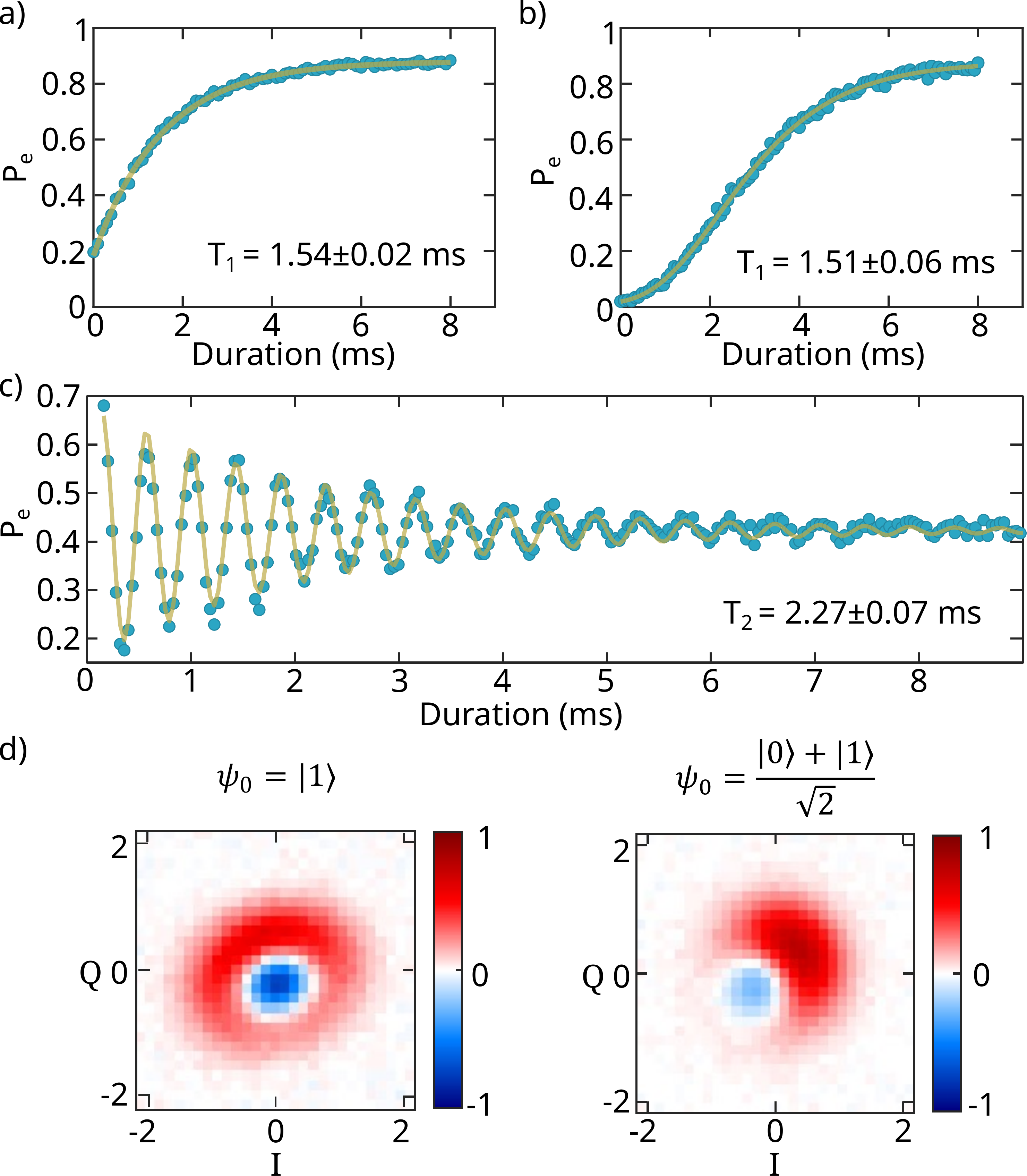}
  \caption{
  $\text{T}_1$ and $\text{T}_2$ measurement of a suspended stripline resonator.
  a) $\text{T}_1$ measurement by preparing the Fock state $\ket{1}$ with SNAP and observing exponential decay.
  b) $\text{T}_1$ measurement by preparing displaced state $\ket{\alpha=\sqrt{2}}$ and observing the double-exponential decay of overlap with $\ket{0}$.
  c) $\text{T}_2$ measurement by preparing an equal superposition of $\ket{0}$ and $\ket{1}$ with SNAP and observing Ramsey-like fringes with a displacement with a variable delay.
  See the supplement for more information on the gate sequence.
  d) Wigner distributions of the initial states as prepared.
  }
  \label{fig:T1T2Data}
\end{figure}

We have measured the $T_1$, $T_2$, and occupation $\bar{n}$ of the storage mode of the resonator, as well as the coherence properties of the transmon ancilla \cite{reagor_quantum_2016}.
The $T_1$ was measured in two different ways (Fig.~\ref{fig:T1T2Data}~a, b).
First, we prepared the Fock state $\ket{1}$ using a selective number-dependent arbitrary phase (SNAP) gate \cite{heeres_cavity_2015, krastanov_universal_2015} and observed its exponential decay to the ground state, using the transmon to read out the resonator's state.
Second, we displaced the cavity to $\alpha=\sqrt{2}$, and observed the double-exponential decay of the population to the ground state as a function of delay time.
We note that both measurements gave similar results.
$T_2$ was measured by preparing $\frac{1}{\sqrt{2}}(\ket{0}+\ket{1})$ using SNAP, waiting a variable amount of time, then displacing with an artificial detuning akin to a $T_2$ Ramsey experiment.
After the transmon readout, we observe decaying oscillations of the population of the $\ket{e}$ state that maps to the $\ket{0}$ state of the cavity (Fig.~\ref{fig:T1T2Data}~c).
For all of the above measurements, we have checked the Wigner functions of the initial states to confirm that our state preparation was working (see Fig.~\ref{fig:T1T2Data} d).
Although our SNAP protocol was not perfect, this only decreases our measurement contrast without affecting the decay times we are measuring.

The results of the measurements are presented in Table~\ref{table:suspended_stripline_results}.
We have measured resonators using multiple materials for the inner and outer conductors: silicon and sapphire center scaffolds, and high-purity (5N) and 6061 alloy aluminum outer conductors.
The silicon scaffolds were 500~\textmu{}m thick, while the sapphire was 100~\textmu{m} \textendash{} this was due to the difficulties associated with lasercutting sapphire, although thicker material would still be possible.
The same transmon was used in all of the measurements.

\begin{table*}
\caption[Measurement results]{Measurement results for several devices with various combinations of center and outer conductors; the same transmon was used every time.
For the resonator, $T_1$ was measured via Fock state ($T_1^\text{F}$) and coherent state ($T_1^\text{C}$) decay; $T_2$ was measured with a Ramsey-liked experiment.
The transmon was measured using standard techniques; it was sometimes not possible to measure $T_2$ due to beating in the transmon g-e transition.
The $\chi/(2\pi)$ was around 700\,kHz for the silicon scaffolds, and around 500\,kHz for the sapphire scaffold; the difference is likely primarily due to the different thicknesses of the dielectrics.
Note that the 5N package needed re-milling and etching due to an unremovable PTFE clamp after its first use. 
The resonator modes were around 5.4\,GHz, the transmon \textendash{} at 6.3\,GHz, and the readout around 8.9\,GHz.
The intervals represent multiple measurements on the same sample.
}
\setlength{\tabcolsep}{5pt}
\centering\begin{tabular}{@{}llcrrrrcrrrr@{}}\toprule
\multicolumn{2}{c}{Device} & \phantom{}& \multicolumn{4}{c}{Resonator} &
\phantom{} & \multicolumn{4}{c}{Transmon}\\
\cmidrule{1-2} \cmidrule{4-7} \cmidrule{9-12}
Center conductor & Al && $T_1^\text{F}$\,(ms) & $T_1^\text{C}$\,(ms) & $T_2$\,(ms) & $\bar{n}$ && $T_1$\,(\textmu{}s) & $T_2$\,(\textmu{}s) & $T_2^\text{E}$\,(\textmu{}s) & $P_e$ \\ \midrule
silicon \#2 & 6061 &&
0.6 & \textemdash{} & \textemdash{} & \textemdash{}&&
42 & 14 & 52 & 5\%\\
silicon \#1 & 6061 &&
0.6-0.8 & 0.5-0.8 & 0.9-1.2 & 0.08 &&
15-31 & \textemdash{} & 15-45 & 0.3\%\\
silicon \#1 & 5N &&
0.6-0.7 & 0.6-0.7 & 0.7-0.9 & 0.03 &&
34-51 & \textemdash{} & 34-68 & 1\% \\
sapphire & 6061 &&
1.3-1.5 & 1.5-1.6 & 2.1-2.4 & 0.05 &&
23-51 & 12-41 & 24-58 & 0.5\% \\
sapphire & 5N &&
0.7 & 0.6-0.8 & 0.9 & 0.08 &&
38-40 & \textemdash{} & 48 & 1.6\% \\
sapphire & 6061 &&
1.0-1.4 & 1.3-1.6 & 0.2 & 0.11 &&
22-31 & 5 & 10-13 & ~4\% \\
\bottomrule
\end{tabular}

\label{table:suspended_stripline_results}
\end{table*}

All of the measured devices had $T_1$ of at least half a millisecond.
The sapphire stripline chip used with the 6061 aluminum package had $T_1$ of over one millisecond. 
The $T_2$ for these measurements was limited by $T_1$ most of the time, except in one case where the transmon's excited state population was anomalously high.
Generally, the $T_2$ was close to the $T_1$ limit, indicating very long dephasing times.
Recalling that the storage mode has almost no participation in the dielectric substrate, one may expect that the silicon and sapphire chips should have the same lifetime.
However, silicon chips were 5 times thicker than the sapphire, and thus had 5 times more sidewall.
As can be seen in the supplement, the sidewall resulting from lasercutting is quite rough, leading us to suspect that it is a source of loss.
This would cause the thicker silicon chips to have lower $T_1$, as observed.
Our data does not enable us to rule out the alternative possibility that the aluminum grown on the two substrates is different.

It may be possible in the future to increase the lifetimes of the resonators by improving the quality of the package and its materials. For example, based on experience with 3D stub cavities, high-purity aluminum has been shown to have lower surface dielectric and conductor losses \cite{reagor_quantum_2016, lei_characterization_2023}.
For this reason, we tested a high-purity aluminum package for our resonator etched using the standard protocol \cite{reagor_quantum_2016}.
However, this device performed worse than expected.
We hypothesize that there is some issue with the particular physical package, either due to contamination or machining imperfections in the surface, and that by trying other copies of the package, we should be able to achieve lifetimes that are several times longer.
Since we attain millisecond coherences with 6061 aluminum in this design, it can be used to avoid 5N aluminum, which is difficult to machine and requires an etch, limiting the possible geometries of the design.

We have demonstrated a new quantum memory design that separates the center conductor from the package, which serves as the outer ground.
This improves modularity by allowing us to replace the pieces individually, as demonstrated in this work.
We have also shown that by concentrating the field in a higher-quality material (the evaporated thin-film on the centerpin), we have relaxed the constraints on the outer conductor, allowing us to achieve lifetimes comparable to stub cavities without using high-purity aluminium or niobium and with no etching.
The absence of a ground plane also means that we can stack several of these centerpin chips in one tunnel, opening up the third dimension for tiling.
Future improvements to this style of resonator can focus on increasing the quality of the sidewalls of the centerpin chip, e.g. by micromachining it out of silicon, stealth dicing, or chemical polishing of the sidewalls.\\

See the supplementary materials for fridge wiring, field profiles of the modes, sidewalls of the lasercut chips, a participation table of the three modes, fabrication recipes, details about an alternative package design including a seam, and the gate sequences used for measurement. \\

We thank Harvey Moseley, Pratheev Sreetharan, and Charles Wehr for aid in mechanical design of an earlier version of the experiment; Yong Sun, Sean Rinehart, Kelley Woods, and Michael Rooks for assistance with device fabrication.
This research was supported by the U.S. Army Research Office (ARO) under grants W911NF-
18-1-0212, and W911NF-23-1-0051. The views and conclusions contained in this document are those of the authors and should not be interpreted as representing official policies, either expressed or implied, of the ARO
or the U.S. Government. The U.S. Government is authorized to reproduce and distribute reprints for
Government purpose notwithstanding any copyright notation herein. Fabrication facilities use was
supported by the Yale Institute for Nanoscience and Quantum Engineering (YINQE) and the Yale
SEAS Cleanroom.

\section*{Author declarations}
\subsection*{Conflict of interest}
R. J. S. and L. F. are founders and shareholders of Quantum Circuits, Inc.
C.U L., S. G., L. K., R. J. S., and L.F. have Patent PCT/US2022/016733 pending.

\subsection*{Author contributions}
Lev Krayzman and Chan U Lei contributed equally to this work.

Concept: L. K., C.U L, S. G., R. S.
Funding acquisition: R. S.
Investigation: L. K., C.U L., J. T.
Fabrication of devices: L. K., C.U L., L. F.
Oversight: R. S.

\section*{Data Availability Statement}

Data are available at reasonable request to the corresponding authors.

\bibliography{BibliographyMaster}

\cleardoublepage
\appendix

\onecolumngrid
\section{Fridge wiring}
\begin{figure*}[hb!]
\centering
\includegraphics[width=0.75\textwidth]{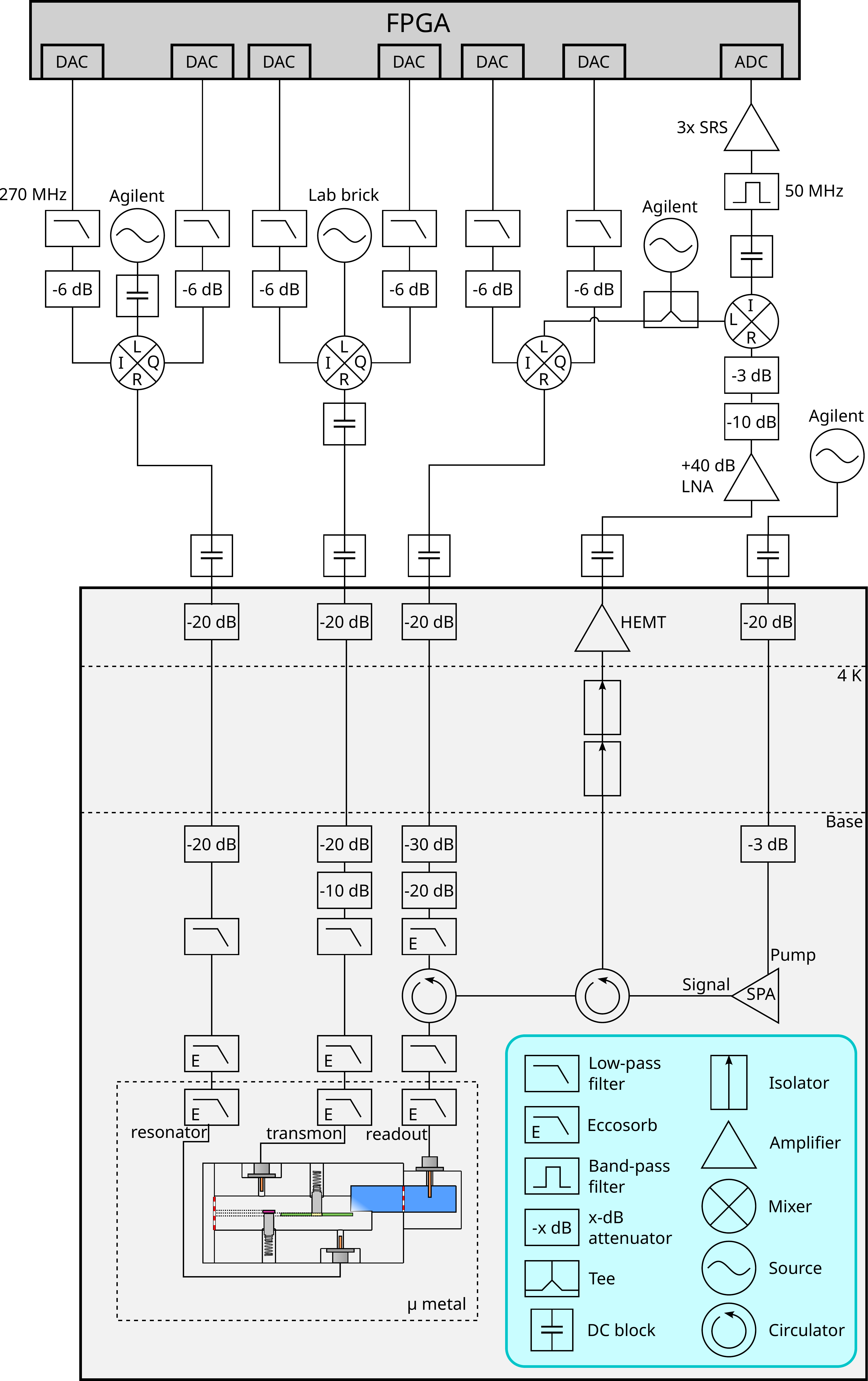}
\caption{Fridge wiring diagram.}
\label{fig:wiring}
\end{figure*}
\cleardoublepage

\section{Field profiles of the modes}
We show the electric field magnitudes of the four relevant modes from the same HFSS simulation as above.
It is plotted on a log (left) and linear (right) scale for each of the modes.\vspace{1.5ex}

\begin{figure*}[h!]
\centering
\includegraphics[width=1\textwidth]{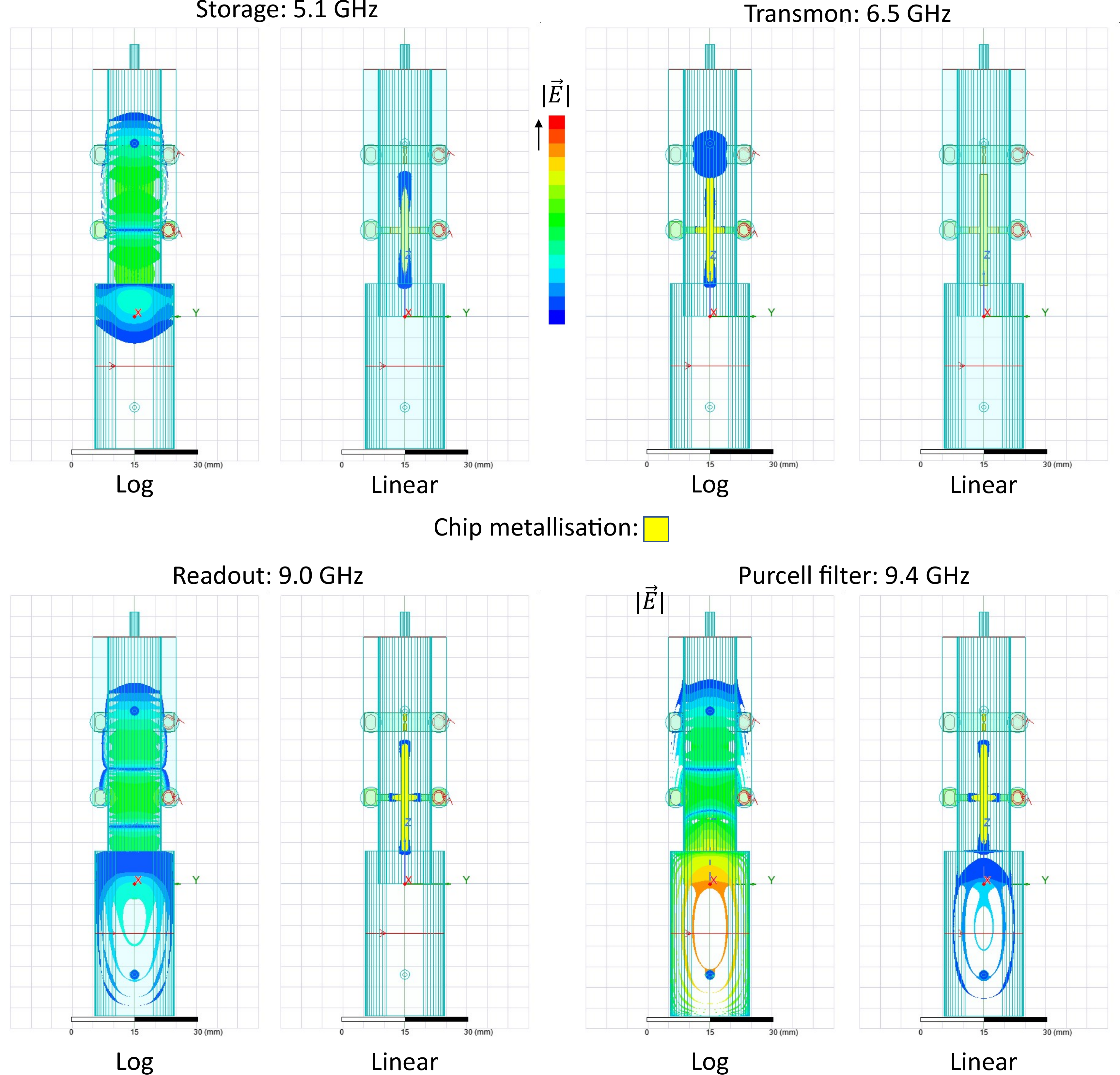}
\caption{Finite-element eigenmode simulations of the electric fields of the four relevant modes in Ansys HFSS.
For each mode, the magnitude of the electric field is presented on a linear and logarithmic scale.
Note that the metallization on the dielectric scaffold is shown in yellow, not to be confused with high electric field.
The storage mode is seen to have a field node at the location of the unmetallized substrate, and to attenuate towards either of the seams.}
\label{fig:mode_simulation}
\end{figure*}
\cleardoublepage

\onecolumngrid
\section{Lasercut chip sidewalls}
\begin{figure*}[h!]
    \centering
    \includegraphics[width=1\textwidth]{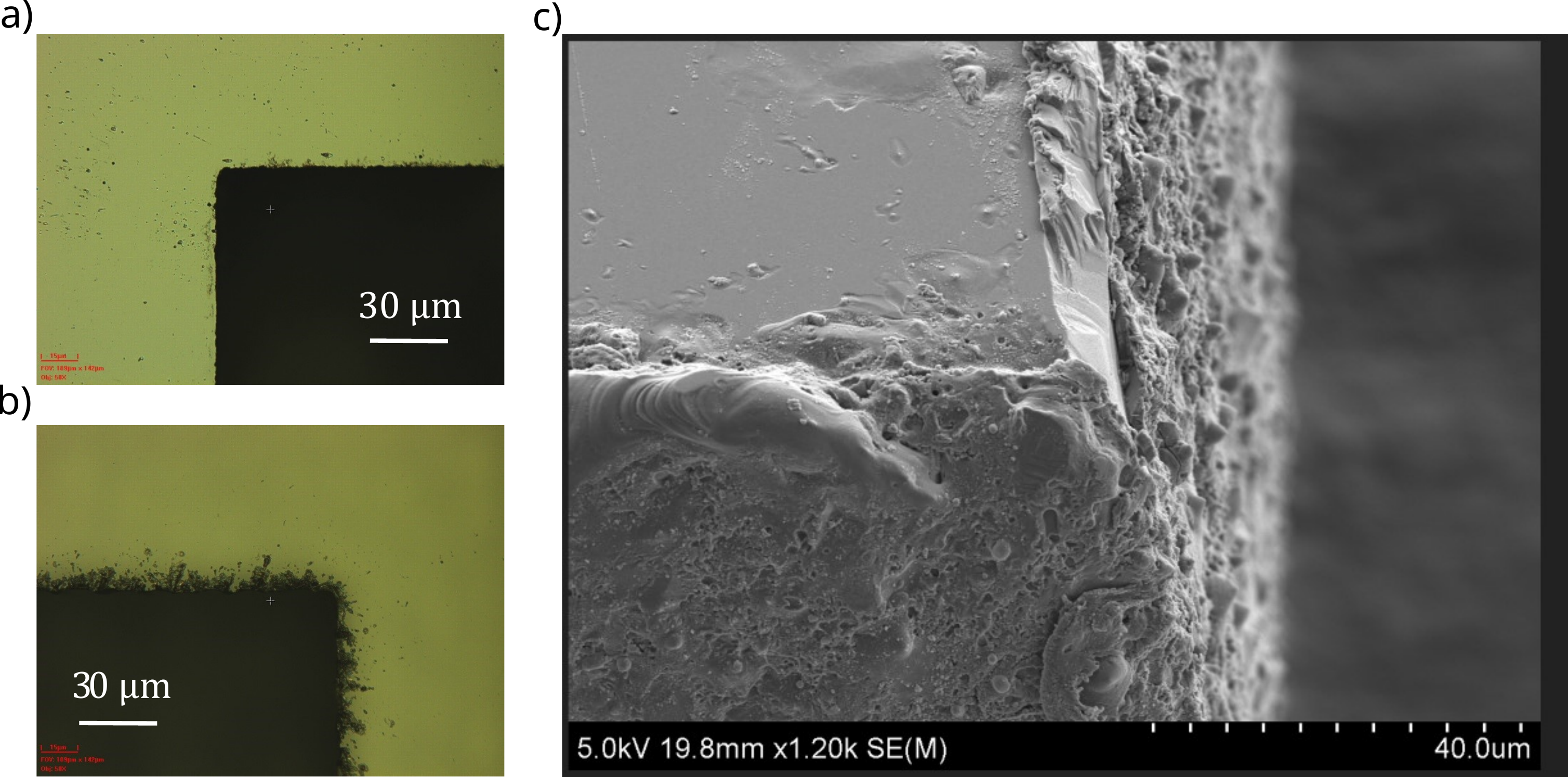}
    \caption{Sidewalls of the lasercut dielectric scaffolds.
    a) Optical micrograph of a lasercut silicon scaffold against a dark background.
    The dots visible on the surface are likely re-deposited material; their density decreases away from the edge.
    b) Optical micrograph of a lasercut sapphire scaffold against a dark background.
    The edge is visibly rougher than for the silicon.
    c) Scanning electron micrograph of a corner of a lasercut silicon scaffold. 
    Note the large number of sharp points microns in size.
    This image is taken after the scaffold has had several hundred nm of aluminum evaporated on it from all sides; the devices do not look qualitatively different prior to evaporation.
    }
    \label{fig:lasercut_chip_sidewalls}
\end{figure*}

\twocolumngrid
\section{Participation table of the three modes}

\begin{table}[h]
\caption{Participation table
}
\setlength{\tabcolsep}{5pt}
\centering\begin{tabular}{llll}
\toprule
 & 
\multicolumn{3}{c}{Participations for mode}\\
\cmidrule{2-4} \\
Region & Storage & Readout & Purcell \\
\midrule \\
Centerpin bulk dielectric & $3\times10^{-5}$ & $2\times10^{-1}$ & $1\times10^{-4}$ \\
PTFE clamp bulk & $2\times10^{-7}$ & $9\times10^{-5}$ & $1\times10^{-6}$ \\
Centerpin conductor & $2\times10^{-5}$ & $2\times10^{-5}$ & $3\times10^{-8}$ \\
Centerpin MA & $1\times10^{-7}$ & $1\times10^{-7}$ & $1\times10^{-10}$ \\
Centerpin SA & $7\times10^{-10}$ & $3\times10^{-6}$ & $2\times10^{-9}$ \\
Package conductor & $4\times10^{-6}$ & $4\times10^{-6}$ & $2\times10^{-5}$ \\
Package MA & $2\times10^{-8}$ & $2\times10^{-8}$ & $6\times10^{-8}$ \\
$y_\text{seam}$ at endcap ($(\Omega \text{m})^{-1}$) & $4\times10^{-10}$ & $8\times10^{-9}$ & $2\times10^{-9}$ \\
$y_\text{seam}$ in Purcell ($(\Omega \text{m})^{-1}$) & $3\times10^{-7}$ & $4\times10^{-6}$ & $3\times10^{-4}$ \\

\bottomrule
\end{tabular}

\label{table:participations}
\end{table}

The device is simulated in Ansys HFSS using the eigenmode solver.
This allows us to directly compute the participations via integrating the fields, as shown in Table~\ref{table:participations}.
We also obtain the Purcell limits of the various modes by placing a 50~$\Omega$ lumped-element resistor in the port of interest and computing the Q of the mode.

\section{Fabrication recipes}

\subsection{Stripline chips}
The wafers were first optionally coated in positive photoresist prior to lasercutting in order to protect them from debris. 
They were then shipped to a company which lasercut them into the cross shapes.
The chips were then solvent cleaned.
Those that had photoresist did have less debris on the surface; we did not perform a study of whether this affected the resulting Q of the final device.

The chips were then inserted into the custom-built rig shown in Fig.~\ref{fig:evaporation}.
The rig was previously prepared with a thin strip of aluminum foil for each wing, which was then wrapped around the cross-beam of the chip.
The purpose of this piece of foil is to create a sharp line between the exposed and shadowed parts of the chip.
The chips were held in place by beryllium copper clamps on each side.
These can be seen on the right wing of Fig.~\ref{fig:evaporation}~b), with the small piece of foil just to the left of the clamp (the clamp appears silvery due to aluminum evaporation onto to it).
Finally, this construction is wrapped in a larger piece of foil, which shadows the remaining parts of the clamping pad, as seen on the left wing in this image.

The rig is angled at 45 degrees, such that when the sample is rotated during evaporation, it is covered from all sides.
After evaporation, an optional capping layer of oxide can be grown in the evaporator by exposing the chip to oxygen.
We have not tested this extensively, but we saw no evidence of this step making any difference in the resulting device.

The stripline chips used in this paper were silicon \#1, with a total evaporation thickness of 600~nm Al on a 500 \textmu{}m-thick silicon chip, silicon \#2, with a total evaporation thickness of 1800~nm Al on a 500 \textmu{}m-thick silicon chip, and sapphire, with a total evaporation thickness of 1000~nm Al on a 100 \textmu{}m-thick sapphire chip.

\subsection{High-purity aluminum etching}
In order to achieve high quality for a device made of high-purity (e.g. 4N, 5N) aluminium, it must be etched \cite{reagor_reaching_2013}.
The current belief is that this either helps by removing mechanical damage caused by the machining, or by modifying the chemical composition of the surface (i.e. removing iron implantations from the tools and in return, adding some phosphorus from the acid \cite{Krayzman_thin-film_2022}).
We use a process based on Ref.~\onlinecite{reagor_reaching_2013,reagor_superconducting_2015}.
First, the cavity is sonicated in solvents in order to remove machining oil, etc.
Then, the tapped holes in the cavity are filled with PTFE screws, to protect the thread without adding contaminants from metal screws.
The cavity is then placed in a PTFE holding rig inside a glass beaker, oriented in such a way as to let bubbles escape.
A magnetic stir bar is placed under the cavity.
The cavity is then etched in Transene Aluminum etchant type A at 50\,$^\circ$C while the liquid is being stirred at around 600 rpm.
When the bath becomes opaque, the cavity is transferred to another bath of etchant; in total, we normally etch in three or four baths.
Each bath takes around 30 minutes.
After the last bath, the cavity is transferred to a beaker of DI water, rinsed for around a minute in the beaker, and then another minute outside the beaker, before being dried with nitrogen.

\onecolumngrid
\subsection{Transmon}
\setlength{\tabcolsep}{6pt}
\begin{table*} [h!]
\caption[Transmon fab]{Transmon fabrication on 100~\textmu{}m-thick sapphire wafers.
The ion mill in the evaporator was done with 3.5:1 Ar:O$_2$ gas, at 250~V beam voltage and 5~mA beam current.
The aluminium evaporation was 20~nm at 20$^\circ$, 30~nm at -20$^\circ$, with 15~min of oxidation at 15~Torr in between.
There was a capping layer of 5~min oxidation at 50~Torr at the end.}
\centering
    \begin{tabular}{l l l}
    \toprule
        Step & Description & Notes \\
        \midrule
        Solvent clean & NMP, acetone, methanol, DI & \\
        Dehydration bake & 5 min at 180~$^\circ$C & 2~min cool\\
        Spin MAA EL13 & 1) 400 rpm, acl = 5, 10 s & \\
         & 2) 2000 rpm, acl = 10, 1:40 & Non-vacuum chuck, used syringe with 0.22~\textmu{}m filter \\
         Bake & 5~min at 180~C & 2~min cool\\
         Spin PMMA A4 & 1) 400 rpm, acl = 5, 10 s & \\
         & 2) 2000 rpm, acl = 10, 1:40 & Non-vacuum chuck, used syringe with 0.22~\textmu{}m filter\\
         Bake & 5~min at 180~$^\circ$C & 2~min cool\\
         Spin PSSA & 3000~rpm, 2~min & Used syringe with 0.22~\textmu{}m filter\\
         Bake & 1~min at 120~$^\circ$C & To remove water \\
         Sputter gold & 90~s & Standard recipe uses 60~s\\
         Write & Raith EBPG 5000+ &  Maximum curent used: 100~nA\\
         Remove PSSA & Rinse with DI for a few seconds & Removes most, but not all, gold\\
         Remove gold & KI for about 20~s & DI rinse \\
         Develop & 3:1 IPA/water at 6~$^\circ$C, 2~min  & Gentle swirl \\
         Deposit aluminium & Evaporate after 30~s ion mill & See caption \\
         Liftoff & NMP at 80~$^\circ$C, 2~hrs & Sonicate after, clean in acetone, methanol, nitrogen dry\\
         Probe & Room-temperature resistances & \\
         Spin dicing resist & SC1827 500~rpm 10~s & \\
         & 1500~rpm 1.5~mins& \\
         Dice &  & \\
         Remove resist & With acetone, methanol & \\
    \bottomrule
    \end{tabular}

\label{table:transmon_fab}
\end{table*}
\cleardoublepage

\twocolumngrid
\section{Alternative package with a seam}

\begin{figure}[ht]
\centering
\includegraphics[width=1\linewidth]{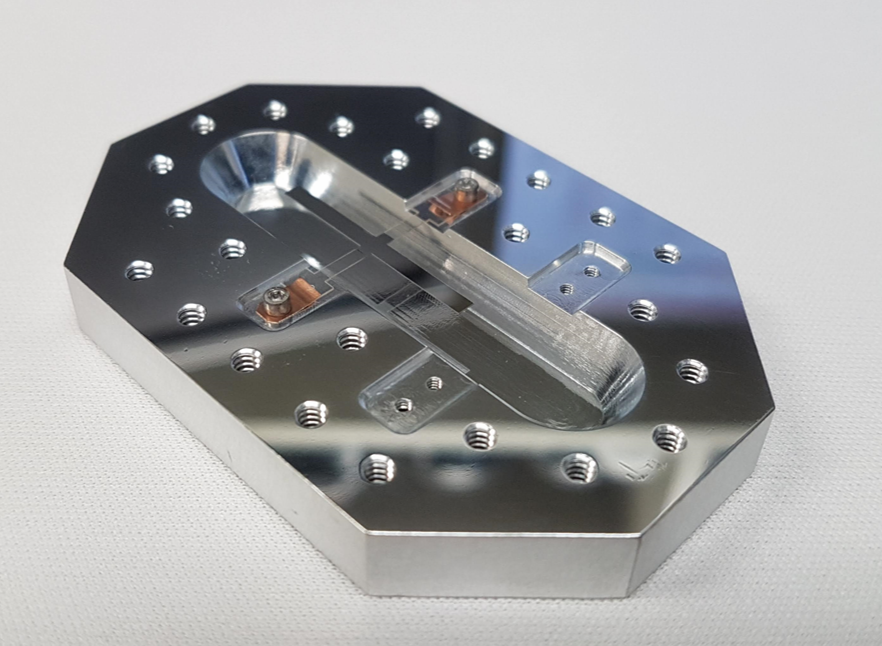}
\caption[Suspended stripline in a seam package]{Photograph of a sapphire stripline suspended in a seam package.
Beryllium copper clamps hold the dielectric scaffold to the package on each side.
The walls of the cavity are machined at an angle, enabling anisotropic metal deposition (e.g. evaporation).
A space is left for a transmon chip.
The top half of the package (not pictured) is almost identical to the bottom.
The one main difference is a coupling port placed directly over the center of the suspended stripline.
The cavity is manually mechanically polished with sandpaper and then Pikal care; the mating surfaces are diamond-turned.
}
\label{fig:suspended_stripline_seam_package}
\end{figure}

A different style of package was also tried with the same suspended stripline centerpins.
In this design, the package was made of two parts, with grooves cut for holding the center conductor and transmon chips.
The chips were held in place with beryllium copper clamps.
The advantage of this design is the relative ease of inserting or removing the chips.
See Fig.~\ref{fig:suspended_stripline_seam_package} for an image of the bottom half of the package with a chip inside \textendash{} the top half (not shown) is very similar.
The sloped walls of the cavity in the package allow us to deposit thin-film materials onto the cavity surface.

One potential concern in this design is the presence of a seam in the package.
The expected $y_\text{seam}$ for the storage mode in this configuration is around $10^{-4}/(\Omega \text{m})$.
With polished or diamond-turned cavity surfaces, we can achieve $g_\text{seam} > 10^4/(\Omega \text{m})$, and with deposited thin-film aluminum, this goes up to $g_\text{seam} >= 10^7/(\Omega\text{m})$ \cite{lei_characterization_2023}.
Therefore, the seam should only become relevant at Q's of hundreds of millions to many billions.
Otherwise, the participations are comparable to that of the traditional stub cavity and the seamless design presented in the main text (Table~\ref{table:suspended_seam_package_participations}).

\begin{table}[h]
\vspace{1ex}
\caption[Participations for seam package for suspended stripline resonator]{Participations for the seam package for a suspended stripline resonator.
The storage and readout modes have almost identical participations, except the readout mode has thousands of times higher $p_\text{diel}$ (here referring to the chip), and a factor of ten higher $y_\text{seam}$.
The participation of the clips is negligible.
}
\centering
    \begin{tabular}{l c c}
    \toprule
        Participation & Storage mode & Readout mode \\
        \midrule
        $p_\text{diel}$ & $4.4\times 10^{-5}$ & $1.2\times10^{-1}$ \\
        $p_\text{cond,stripline}$ & $3.0\times10^{-5}$ & $3.1\times10^{-5}$ \\
        $p_\text{cond,package}$ & $6.2\times10^{-6}$ & $5.7\times10^{-6}$ \\
        $p_\text{MA,stripline}$ & $2.1\times10^{-7}$ & $1.7\times10^{-7}$ \\ 
        $p_\text{MA,package}$ & $2.7\times10^{-8}$ & $3.4\times10^{-8}$ \\ 
        $y_\text{seam}$ & $1.1\times10^{-4}/(\Omega \text{m})$ & $1.4\times10^{-3}/(\Omega \text{m})$ \\
    \bottomrule
    \end{tabular}
\label{table:suspended_seam_package_participations}
\end{table}

We have measured several centerpins in two packages of this design.
Both packages were made of 6061 aluminum, and had either 200 or 400~nm of aluminum evaporated onto them.
The centerpin chips were made of either silicon or sapphire with different amounts of aluminum evaporated onto them; most also had a capping oxide layer grown in the evaporator.
See Table~\ref{table:suspended_seam_package_measurements} for a summary of the devices measured and measurement results.

Although some devices had reasonably high Q's, a large jitter of the resonance was observed in a number of cooldowns.
This could be observed directly on a VNA in a spectroscopy measurement, with the resonance moving many linewidths during a single sweep.
We were unable to remove this effect in this package design.

\begin{table*} 
\caption[Measurements of suspended striplines in seam packages]{Measurements of suspended striplines in seam packages in reflection and ringdown.
Two packages and several chips were used.
All but the first two chips received a ``capping'' oxidation step in the aluminium evaporator in order to have more controlled oxide rather than native growth.
One of the chips was not rotated during evaporation.
All devices were measured in the frequency domain with microwave spectroscopy, three were additionally measured in time-domain with ringdown.
Many of the devices demonstrated large amounts of jitter on the VNA, indicating some sort of shaking.
This tended to improve (but not necessarily disappear entirely) with the pulse tube of the dilution refrigerator turned off.}
\centering
    \begin{tabular}{l l l r c l}
    \toprule
        Package & Stripline chip & Stripline metallisation & $Q_{i,\text{spectroscopy}}$ & $Q_{i,\text{ringdown}}$ & Jitter \\
        \midrule
        400\,nm & 500\,\textmu{}m Si & 2\,\textmu{}m Al, no oxidation & $44\times10^6$ & \textemdash{} & No \\
        400\,nm & 500\,\textmu{}m Si & 2\,\textmu{}m Al, no oxidation & $48\times10^6$ & \textemdash{} & No \\
        200\,nm & 500\,\textmu{}m Si & 600\,nm Al with oxidation & $37\times10^6$ & \textemdash{} & Yes\\
        400\,nm & 500\,\textmu{}m Si & 800\,nm Al with oxidation & $37\times10^3$ & \textemdash{} & No \\
        400\,nm & 100\,\textmu{}m sapphire & 600\,nm with oxidation & $108\times10^6$ & $140\times10^6$ & Yes \\
        200\,nm & 100\,\textmu{}m sapphire & 600\,nm with oxidation & $127\times10^6$ & $145\times10^6$ & Yes \\
        200\,nm & 100\,\textmu{}m sapphire & 600\,nm Al no rotation, with oxidation & $9.6\times10^6$  & \textemdash{} & Yes \\
        200\,nm & 100\,\textmu{}m sapphire & 800\,nm Al with oxidation & $37\times10^6$ & $33\times10^6$ & No \\
    \bottomrule
    \end{tabular}
\label{table:suspended_seam_package_measurements}
\end{table*}

\twocolumngrid
\section{Gate sequences}
\begin{figure}[h]
\centering
\includegraphics[width=1\linewidth]{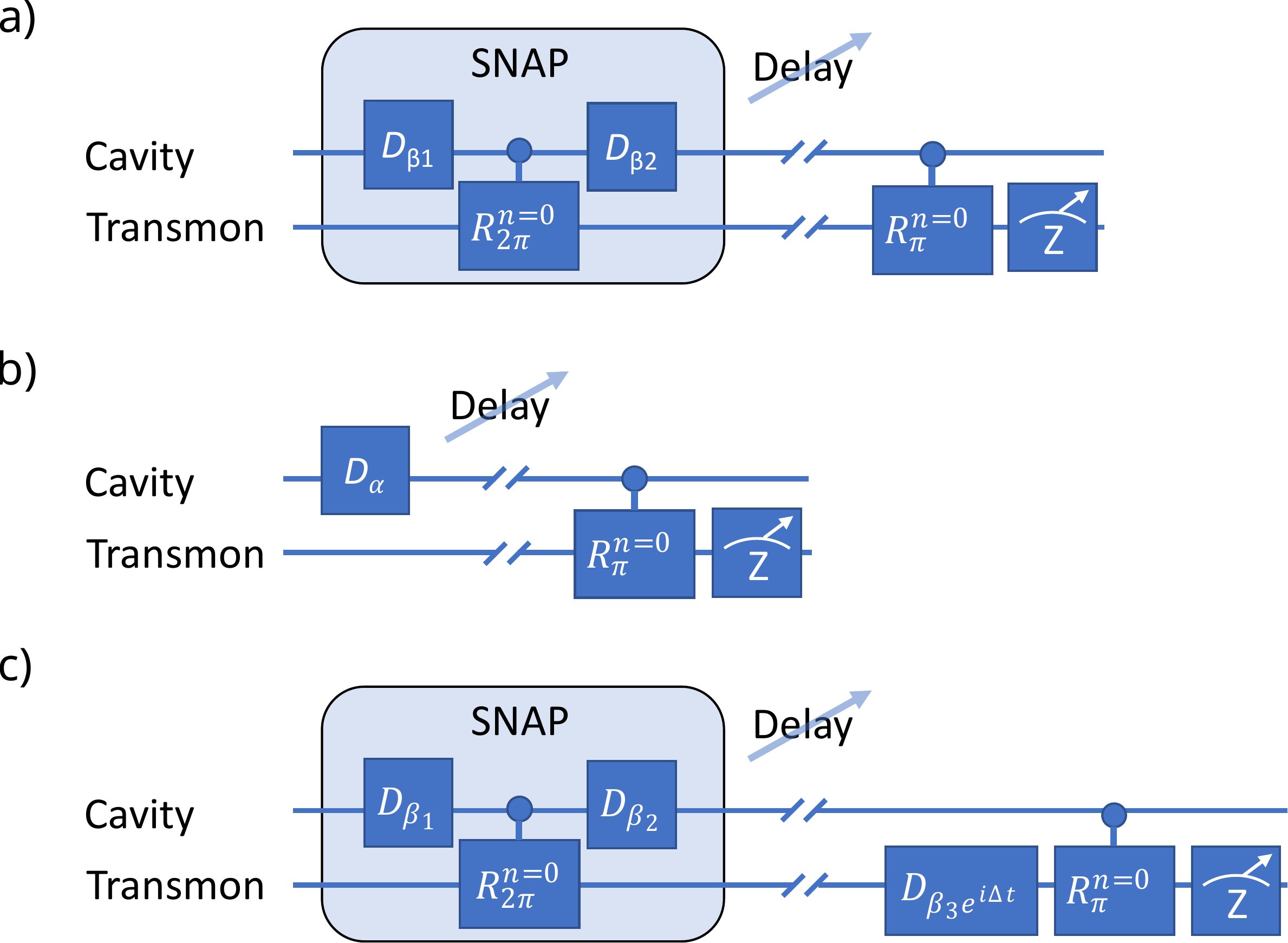}
\caption{Gate sequences used for the T1 (a, b) and T2 (c) measurements.
a): Prepare $\ket{1}$ with SNAP, wait variable delay, readout using $\pi$ pulse selective on 0 photons.
b): Displace cavity to $\ket{\alpha}$, wait, readout.
c): Prepare $\frac{\ket{0} + \ket{1}}{\sqrt{2}}$ with SNAP, wait, displace back with variable phase, readout.
The values used for the displacements and rotations are taken from \cite{reagor_quantum_2016}.}
\label{fig:gate_sequences}
\end{figure}

\end{document}